\documentclass[twocolumn,pra,a4paper,floatfix]{revtex4}
\usepackage{amsmath}
\usepackage{amssymb}
\usepackage{amsfonts}

\usepackage{fancybox}
\usepackage{eepic}
\usepackage{times}
\usepackage{latexsym}
\usepackage{pifont}
\usepackage{epsf}
\usepackage{pstricks}
\usepackage{epsfig}

\newcommand{\bi}[1]{\mbox{\boldmath $#1$}}

\newcommand{\opencircle}{\mbox{\Large$\circ\,$}}  

\newcommand{\fullcircle}{\mbox{{\Large$\bullet\,$}}} 
\newcommand{\fullsquare}{\,\vrule height5pt depth0pt width5pt}
\newcommand{\dotted}{\protect\mbox{${\mathinner{\cdotp\cdotp\cdotp\cdotp\cdotp\cdotp}}$}}
\newcommand{\dashed}{\protect\mbox{-\; -\; -\; -}}

\newcommand{\full}{\protect\mbox{------}}
\newcommand{\centre}[2]{\multispan{#1}{\hfill #2\hfill}}

\begin{document}

\title{A discrete time-dependent method for metastable atoms
       in intense fields} 

\author{Liang-You Peng$^{\ast,\dag}$, J F McCann$^{\dag}$, Daniel Dundas$^{\dag}$,
 K T Taylor$^{\dag}$ and I D Williams$^{\dag}$}

\affiliation{$^\ast $International Research Centre for Experimental Physics, \\
         $^\dag $School of Mathematics and Physics, \\
           Queen's University Belfast, \\
           Belfast BT7 1NN, Northern Ireland, UK.}

\begin{abstract}
 The full-dimensional time-dependent Schr\"{o}dinger equation for the electronic
 dynamics of  single-electron systems in intense external fields is 
 solved directly using a discrete method. 
 Our approach combines  the finite-difference  and 
 Lagrange mesh methods. The method is applied to calculate the quasienergies 
 and ionization probabilities of atomic and molecular systems in intense
  static and dynamic electric fields. The gauge invariance and accuracy of 
  the method  is established. Applications to multiphoton ionization  of 
  positronium and hydrogen atoms and molecules are presented. At very high
   intensity above saturation threshold, we extend the method 
   using a scaling technique to estimate the quasienergies of metastable 
   states of the hydrogen molecular ion. The results are in good agreement 
   with recent experiments.   
 \end{abstract}

\maketitle

\begin{section}{Introduction}

Matter exposed to intense laser fields has attracted  extensive research in 
the past two decades. Due to the competition  between external forces 
induced by the laser field and Coulomb interactions, which bind the system 
together, the nature of the metastable states that arise is strongly 
controlled by the nature of the external field and is not simply 
a feature of the atomic or molecular system. In such circumstances 
the internal and external fields are on an equal footing and 
not separable. 
   
The study of these effects in few-electron atoms   have   become 
accessible recently both experimentally and theoretically because 
of advances in experimental techniques and the availability of 
supercomputers. Compared with atoms, molecules
in intense laser fields are much more complicated not only because of the 
multi-center nature of the problem but also due to the additional 
vibrational and rotational degrees of freedom associated with the 
nuclei~\cite{Codling,Giusti}. In polyatomic systems, the energy transfer 
to collective motion leads to extremely high-energy secondary 
photons, ions and electrons~\cite{Posthumus,Krainov}.

Standard perturbation theory is no longer applicable when the forces 
induced by the applied laser field are comparable to the binding forces of 
the system. If the external field is periodic then the system 
is metastable and has a well-defined quasienergy spectrum. Suppose 
the unperturbed ground-state has a real energy $E_i$, then 
the effect of the external field is to produce an energy 
shift $\Delta$ and decay width rate $\Gamma/\hbar$, so that 
the quasienergy  has the usual form 
$E=E_i+\Delta -i\Gamma/2 $. 
This describes a steady-state decay and can be treated very effectively 
with time-independent methods such as the Floquet method.  However, 
at the very highest  intensities 
the transient aperiodic effects of short duration pulses of 
dynamic fields mean that the shift and rate are poorly defined. Under these 
conditions a time-dependent approach is essential. In this paper 
we describe such an approach applied to the regimes where 
the quasienegy is well-defined and to time-dependent problems where 
this is not the case. We find the method works efficiently and accurately 
for both cases, and thus is well-suited to the study of metastable states 
under all conditions.  Recent advances in experimental technology in the 
utilisation of 
high intensity lasers have led to the creation 
and study of short-lived atomic and molecular states. In this paper 
we apply the method to problems of this type, though the method 
is of more general applicability.
 With very short pulses, not only the laser 
envelope but even the phase of the field can be  
important~\cite{Paulus}. Therefore, one has to directly integrate  the 
full-dimensional Schr\"{o}dinger equation in order to describe accurately 
the underlying physical mechanisms~\cite{Chelkowski1,Symth,Dundas2}  and 
obtain data relevant to experiments.
 
Although one-dimensional models have been routinely used in describing 
atoms and molecules in strong laser fields~\cite{Steeg}, {\it ab initio}  
full-dimensional quantum-mechanical  calculations are very important for 
exactly calculable few-electron systems. As discussed in~\cite{Steeg}, 
the numerical results for excitation, dissociation and ionization  in  
simplified models  are strongly sensitive to the parameters chosen. Full-dimensionality calculations 
 without approximation
 are really necessary to establish accurate data  
 as  benchmarks to assess the quality or regime of applicability of 
 other calculations and to  produce results 
 which are comparable with  experimental observations.
 
 We have recently developed  a very accurate and efficient numerical method to
 study metastable states in high intensity fields. This was applied to 
 the fragmentation of the hydrogen molecular
  ion in an intense laser field beyond the Born-Oppenheimer 
 approximation~\cite{Dundas2,Dundas3,Peng}. In this paper, we make 
 a detailed investigation of  this method applied to metastable 
 atomic and molecular systems.
 We discuss the variational characteristics of this method and confirm
  its accuracy, convergence and gauge invariance.  
 Results are compared to other theoretical estimates of the 
 metastable states, and we compare with experimental results for 
 the ionization of the hydrogen molecular ion at high intensities.  
\end{section}

\begin{section}{Discrete Time-Dependent Schr\"odinger Equation}
Our goal is the direct solution of the time-dependent Schr\"odinger equation 
of an arbitrary one-electron systems in strong external electric fields
\begin{equation}
H_{\rm e} \Psi_e( \bi{r},t) = i \hbar {\frac{\partial}{\partial t}} 
\Psi_e( \bi{r},t),
\end{equation}
where
\begin{equation}
H_{\rm e} \equiv \frac{1}{2 m_e} (\bi{p} +e \bi{A}(\bi{r},t) )^2 +
V(\bi{r},t),
\end{equation}
and $\bi{A}(\bi{r},t)$  is the (external) vector potential; 
$ V(\bi{r},t)$ is the (internal) scalar potential. 
An example of such a problem is a one-electron molecular system in 
an intense laser field; the internal   field arising from 
the positively-charged nuclei. In order to simplify the model 
and to understand the electronic dynamics 
in isolation, we consider the hydrogen molecular ion with 
the nuclei fixed in space. The nuclei, labelled 1 and 2, are a 
fixed distance $R$ apart, and have  charges $Z_1$ and $Z_2$. 
The origin of the coordinate $\bi{r}$  
is located at the internuclear midpoint, with the 
electronic Hamiltonian takes the form
\begin{eqnarray}
H_{\rm e} &\equiv& \frac{1}{2 m_e} (\bi{p} +e \bi{A}(\bi{r},t) )^2 \nonumber \\
&+&
\frac{e^2}{4\pi \epsilon_0}\left( 
-\frac{Z_1}{r_1} -\frac{Z_2 }{r_2} + \frac{Z_1Z_2}{R}
\right),
\end{eqnarray}
with $r_1 = | \bi{r}+\textstyle\frac{1}{2} \bi{R} | $
and  $r_2 = | \bi{r}-\textstyle\frac{1}{2} \bi{R} | $, and 
$ \bi{A}$ the vector potential. We include the constant 
internuclear potential for reasons of convention. 
Removing the quadratic term $(e^2/2 m_e) \bi{A}^2$ by 
gauge transformation,  and making the dipole approximation, the 
Hamiltonian in the Coulomb gauge can be written as
\begin{eqnarray}
H^{(V)}_{\rm e} &\equiv& \frac{1}{2 m_e} \bi{p}^2 + \frac{e}{m_e} \bi{A}(t) 
\cdot \bi{p}  \nonumber \\
&+&\frac{e^2}{4\pi \epsilon_0}\left( 
-\frac{Z_1}{r_1} -\frac{Z_2}{r_2} + \frac{Z_1Z_2}{R} \right),
\end{eqnarray}
while in the length gauge: $\bi{E}(t) = -\ \partial \bi{A} /\partial t$
\begin{eqnarray}
H^{(L)}_{\rm e} &\equiv& \frac{1}{2 m_e} \bi{p}^2 +e\bi{r} \cdot \bi{E}(t) \nonumber \\ 
&+&\frac{e^2}{4\pi \epsilon_0}\left(
-\frac{Z_1}{r_1} -\frac{Z_2}{r_2} + \frac{Z_1Z_2}{R}
\right).
\end{eqnarray}
Monochromatic light with 
linear polarization  parallel to the internuclear axis implies 
a cylindrical symmetry about this  axis. 
Associated with this symmetry is a good quantum number, $\Lambda$, proportional to 
the projection of angular  momentum along the  axis. 
Thus the  electron position can  be completely described by the radial,
 $\rho$, and axial, $z$, coordinates
with respect to an origin taken at the midpoint between the nuclei. 
The time-dependent Schr\"odinger equation 
reduces to  a 2+1 dimensional partial differential equation. 
Hence in atomic units
\begin{eqnarray}
H^{(L)}_{\rm e}(R;\rho,z;t) =& -&\!\!\frac{1}{2} 
    \left( \frac{\partial^2}{\partial z^2} +\frac{\partial^2} {\partial \rho^2} + \frac{1}{\rho}
   \frac{\partial}{\partial \rho}\right)\nonumber \\
   & + & V_{\rm m}(R,\rho,z) + V_{\rm ml}^{(L)}(z,t),  
\end{eqnarray}
where the electronic potential energy is the non-separable singular function
\begin{eqnarray} 
     V_{\rm m}(R,\rho,z)=& -&\frac{Z_1}{
     \sqrt{\rho^2+(z+ \textstyle\frac{1}{2} R)^2}} \nonumber \\
     &-&\frac{Z_2}{\sqrt{\rho^2+(z-\textstyle\frac{1}{2} R)^2}}\nonumber \\ 
     &+& \!\!\frac{\Lambda^2}{2 \rho^2} +\frac{Z_1Z_2}{R},
\end{eqnarray}  
and the molecule-laser interaction term is
\begin{equation}
   V^{(L)}_{\rm ml}(z,t)= z E(t). 
\end{equation}

Consider a short  optical pulse that is approximately 
monochromatic, within the bandwidth limit, with a well-defined 
peak intensity. This can be simulated 
by choosing an electric field  of the following form
\begin{equation}
   E(t)=E_0 f(t) \cos \omega_L t,
\end{equation}
where the pulse envelope, $f(t)$, is given by
\begin{equation}
   f(t)= \left\{
   \begin{array}{lc}  
      \frac{1}{2}\left[1-\cos\left(\frac{\pi t}{\tau_1}\right)\right] & 
      0\leq t\leq \tau_1\\
      1 & \tau_1\leq t \leq \tau_1+\tau_2 \\
      \frac{1}{2}\left[1-\cos\left(\frac{\pi(t-\tau_2-
      2\tau_1)}{\tau_1}\right)\right] & 
      \tau_1+\tau_2 \leq
      t\leq \tau_2+2\tau_1\\
      0 & t<0, t> \tau_2+2\tau_1
  \end{array}
  \right.
\end{equation}
and in which $E_0$ is the peak electric field, the pulse ramp time is $\tau_1$ 
and the pulse duration $\tau_2$, with associated bandwidth 
$\Delta \omega= 1/\tau_2 $. Since the peak field, $E_0$, is related to the 
cycle-average intensity, $\langle I \rangle $,  by the relation
$\langle I \rangle = {\textstyle \frac{1}{2}}c\epsilon_0  E_0^2$
then the conversion formula is 
$E_0 \approx  5.338 \times  10^{-9} \sqrt{\langle I \rangle}$ if 
the intensity $\langle I \rangle$ is  in W cm$^{-2}$, and the 
field strength in atomic units. The corresponding ponderomotive energy is 
$U_P=E_0^2/(4\omega_L^2)$. Consider the molecule initially in 
the ground state $X\ ^2\Sigma_g^+$. It is convenient to change the dependent 
variable to remove the first-derivative in $\rho$ as follows
\begin{equation}
   \phi(\rho,z,t)= (2\pi \rho)^{1/2} \psi(\rho,z,t),
\end{equation}
so that the time-dependent equation is 
\begin{eqnarray}
i\frac{\partial }{\partial t} \phi(\rho,z,t)  &=& 
\left[ T_z +T_{\rho} + V_{\rm m}(\rho,z,R)\right. \nonumber\\
&&+ \left. V_{\rm ml}^{(L)} (z,t) \right] \phi(\rho,z,t) 
, 
\label{eqn:tde}
\end{eqnarray}
where 
\begin{equation}
    T_{\rho} \equiv -\frac{1}{2} \left(\frac{\partial^2}{\partial \rho^2} + 
    \frac{1}{4\rho^2}\right),
    \ \ \ \ \ \ \ \ \ \ \   
      T_{z} \equiv -\frac{1}{2} \left(\frac{\partial^2} {\partial z^2} \right),
\end{equation} 
with the normalization convention
\begin{equation}
    \int_{0}^{\infty}d\rho\int_{-\infty}^{+\infty}dz
    \left|\phi(\rho,z,0)\right|^2  =  1.
\end{equation}
This 2+1 dimensional equation, given in Eq.~(\ref{eqn:tde}), can be 
discretized on an $N_\rho \times N_z \times N_t$ space-time grid. 
We label the $N_{ \rho}$ radial nodes by, 
$\{\rho_1, \rho_2 , \dots \rho_{i},\dots \rho_{N_{\rho}} \}$,
while the $N_z$ axial grid points are denoted by, 
$\{z_1, z_2 , \dots z_{j},\dots z_{N_z} \}$. The evolution progresses through 
the sequence of times $\{t_1, t_2 , \dots t_{k},\dots t_{N_t} \}$. Thus the 
wavefunction can be written as the array 
$\phi(\rho,z,t) \rightarrow \phi(\rho_j,z_i,t_k) \equiv \phi_{ijk}$. The method 
of discretization of the Hamilton divides the axial and radial coordinates into 
subspaces.  Two distinct but complementary grid methods are used for the 
subspaces. The radial subspace is discretized over a semi-infinite range using 
a small number $N_{\rho}$ of unevenly spaced points that are the nodes of 
global interpolating functions that enter the Lagrange mesh technique. 
This leads to a small dense matrix for the Hamiltonian in the 
$\rho$-subspace. On the other hand the axial coordinate subspace is 
represented by a large number of equally-spaced  points that are the 
mesh points of a finite-difference scheme. The associated 
subspace Hamiltonian matrix is large but sparse. Our approach is
tailored to the requirements of accuracy and computational efficiency.

\begin{subsection}{Discretization  of the Hamiltonian in the radial subspace }
For wavefunction character in the $\rho$-coordinate we have found that a 
Lagrange mesh~\cite{Dundas2,Baye1} can provide a more efficient 
discretization scheme compared with that provided by finite-difference 
formulae. This type of grid can be chosen to accommodate 
short-range singularities or long-range behaviour, and can be scaled 
in length or number of grid points to improve accuracy with a modest number of points. 
We rescale the radial variable as follows, $\rho=h x$ where $h$ is
 some arbitrary scaling factor, and  $0 \leq x < +\infty$.
Then consider the set of functions
\begin{equation}
    \varphi_n(x) =
    \left( \frac{n!}{\Gamma(\alpha+n+1)}\right)^{1/2} 
    x^{\alpha/2}e^{-x/2}L_n^{(\alpha)}(x),
\end{equation}
where $L_n^{(\alpha)}(x)$ are  the generalized Laguerre polynomials
\begin{equation}
L_n^{(\alpha)}(x) \equiv \frac{1}{n!}\  e^x x^{-\alpha}\ 
\frac{d^n}{dx^n}( e^{-x} x^{\alpha+n}).
\end{equation} 
These functions form an  orthonormal set on the domain\newline $0 \leq x < \infty$
\begin{equation}
    \int_0^{\infty} \varphi_m(x)\; \varphi_n(x) \; dx  = \delta_{mn}. 
\end{equation}
Then, for any given value of $\alpha$, one can construct an $N_{\rho}$-point grid 
based upon Gauss quadrature rules. Choosing the grid points ($x_1,x_2,\dots,x_{N_{\rho}}$)  
as the $N_{\rho}$ solutions of $ L_{N_{\rho}}^{(\alpha)}(x)=0$,  the quadrature weights 
corresponding to these pivots are given by the Christoffel numbers $\lambda_i$, where
\begin{equation}
    \lambda_i^{-1} = x_i{\varphi'_{N_{\rho}}}(x_i)^2.
\end{equation} 
One can define the set of differentiable functions 
\begin{equation}
f_i(x)= \lambda_i^{-1/2}\left(\frac{1}{{\varphi'_{N_{\rho}}}(x_i)}\right)\frac
{{\varphi_{N_{\rho}}}(x)}{x-x_i},
\end{equation}
where these mesh functions have  properties both of Lagrange interpolation functions 
and exact discrete orthogonality, that can be summarised as follows
\begin{eqnarray}
\lambda_i^{1/2}f_i(x_j) &= & \delta_{ij}, \\
\sum_{k=1}^{N_{\rho}}  \lambda_i f_i(x_k) f_j(x_k) & =& \delta_{ij}. 
\end{eqnarray}
The matrix element of an operator or function 
$Q(x,\frac{\partial}{\partial x})$, in this basis is given by
\begin{eqnarray}
&&\int_0^{\infty} f^*_i(x)  Q(x,\frac{\partial}{\partial x}) f_j(x) dx = \nonumber \\
& &  \lambda_i^{-1/2}  \left[ Q(x,\frac{\partial}{\partial x}) 
f_j(x) \right]_{x_i}  +\varepsilon 
\end{eqnarray}
where the error, $\varepsilon$, in the Gaussian quadrature depends on ${N_{\rho}}$, $\alpha$ 
and the form of $Q$. Expanding the wave function in this basis
\begin{equation}
    \phi(\rho,z,t) = \sum\limits_{i=1}^{N_{\rho}}(\lambda_i)^{1/2} \phi(h x_i,z,t)
     f_i(x),   
\end{equation}
In the radial subspace the kinetic energy is represented by the dense 
matrix~\cite{Baye1}
\begin{equation}
     (T_{\rho})_{il} =  \left\{
   \begin{array}{lc}  
     \displaystyle \frac{1}{2h^2} \left( \frac{(\alpha+1)^2}{4x_i^2}+ S_{ii} 
     \right), & i=l\\
    \displaystyle  \frac{(-1)^{i-l}}{2h^2} \left[\frac{1}{2} 
    \frac{\alpha+1}{\sqrt{x_ix_l}}\left(\frac{1}{x_i} +
     \frac{1}{x_l} \right)+ S_{il}\right],& i\not=l  
   \end{array}
   \right.
\end{equation}
with 
\begin{equation}
     S_{il} = (x_ix_l)^{1/2} 
     \sum\limits_{k\not=i,l}x_k^{-1}(x_k-x_i)^{-1}(x_k-x_l)^{-1}.
\end{equation}
Then the  Hamiltonian in the radial subspace has the form
\begin{eqnarray}
(H_z)_{il} &\equiv&  \delta_{il} (T_z) +(T_{\rho})_{il} 
    + \delta_{il} V_{\rm m}(\rho_i,z,R) \nonumber \\
    &+&\delta_{il} V_{\rm ml} (z,t).
\end{eqnarray}
The last grid point, the largest root of $L_{N_{\rho}}^{(\alpha)}(\rho/h)=0$, 
defines $\rho_{\rm max}$, the radius of the cylindrical box. 
\end{subsection}
\begin{subsection}{Discretization of the Hamiltonian in the axial subspace} 
The axial coordinate grid is chosen to be a set of $N_z$ equally spaced 
points which cover the range $-z_{\rm max} \leq z \leq
z_{\rm max} $ with a separation $\Delta z= 2z_{\rm max}/(N_z-1) $, so that 
$z_j = -z_{\rm max} +(j-1) \Delta z$.  We choose the method 
of finite differences to treat this coordinate in order to 
make effective use of parallel processing and for appropriate 
treatment of wavefunction dependence on $z$~\cite{Dundas1}. The sparsity of the 
matrix and confinement of communication to that between nearest neighbours 
is ideal for efficient calculation. 
For example, the kinetic energy 
can be evaluated by the five-point finite central
difference formula
\begin{eqnarray}
 (T_z)_{jm} = \frac{1}{24 (\Delta z)^2}
(&&\!\!\!\!\!\!\delta_{j+2,m} -16 \delta_{j+1,m} +30 \delta_{j m} \nonumber \\
 &&\!\!\!\!\!\!- 16 \delta_{j-1,m} +\delta_{j-2, m} ),
\end{eqnarray}
with error proportional to $(\Delta z)^4$ and resulting in 
a sparse (pentidiagnonal) matrix. The momentum operator arises 
in the velocity gauge perturbation and is given by
\begin{eqnarray}
 \left[ V^{(V)}_{\rm ml} \right]_{jm} = \frac{-iA(\rho_i,t)}{12 (\Delta z)}
(&-&\delta_{j+2,m} +8 \delta_{j+1,m}\nonumber \\
 &-& 8\delta_{j-1,m} +\delta_{j-2, m} ),
\end{eqnarray}
with error proportional to $(\Delta z)^5$.

Therefore  the augmented 
$(N_{\rho} \times  N_{z}) \times ( N_{\rho} \times  N_{z}) $ Hamiltonian matrix 
takes the form
\begin{eqnarray}
\left[ H_e^{(L)} (t) \right]_{il,jm} &\equiv&  
\delta_{il}\ (T_z)_{jm} +(T_{\rho})_{il}\ \delta_{jm} \nonumber \\
    &+& \delta_{il} \delta_{jm} V_{\rm m}(\rho_i,z_j,R) \nonumber \\
    &+&
    \delta_{il} \delta_{jm} V^{(L)}_{\rm ml} (z_j,t),
\end{eqnarray}
in the length gauge and
\begin{eqnarray}
\left[ H_e^{(V)} (t) \right]_{il,jm} &\equiv & 
\delta_{il}\ (T_z)_{jm} +(T_{\rho})_{il}\ \delta_{jm} \nonumber \\
    &+& \delta_{il} \delta_{jm} V_{\rm m}(\rho_i,z_j,R) \nonumber \\
    &+&\delta_{il}  \left[ V^{(V)}_{\rm ml} (z_j,t)\right]_{jm},
\end{eqnarray}
in the velocity gauge.
\end{subsection}
\begin{subsection}{Discretization and propagation in time}
The wavefunction is discretized on a grid so that it makes up a vector 
of $N_\rho \times N_z$ components. At time $t_k$ the $(i,j)$th element
of the vector ${\bi{\sf v}}_k$ can be defined as
\begin{equation}
    (\bi{\sf v}_k)_{i,j} = \phi(hx_i,z_j,t_k),
\end{equation}
while the  $(il,jm)$th element of the matrix $\bi{\sf H}_k$ is defined
\begin{equation}
    ({\bi{\sf H}}_k)_{il,jm} =  \left[H(t_k) \right]_{il,jm}, 
\end{equation}
so that the time evolution is described by the equation
\begin{equation}
 \bi{{\sf H}}(t) \bi{\sf v}(t) = i \dot{\bi{\sf v}}(t).
\end{equation}
Suppose that time is divided  so that $t_{k+1} \equiv t_k + \Delta t$
then the solution can be propagated using the unitary evolution matrix 
\begin{equation}
\bi{\sf v}_{k+1} \equiv  \bi{\sf U}(t_k + \Delta t, t_k)\bi{\sf v}_k \approx \exp(-i \bi{\sf H}_k\;\Delta t) \bi{\sf v}_k. 
\end{equation}
Evaluation of this exponential is carried out using a Krylov subspace
decomposition~\cite{Symth,Dundas2}. Using the Arnoldi 
algorithm~\cite{arnoldi:1951} we construct an orthonormal set 
of vectors, $\left[\bi{\sf q}_0, \bi{\sf q}_1, \bi{\sf q}_2, \dots, \bi{\sf q}_{n_a}\right]$, which span the
Krylov subspace 
\begin{equation}
K_{n_a}(\bi{\sf H}_k, \bi{\sf v}_k) \equiv \mbox{span } \left\{\bi{\sf v}_k, \bi{\sf H}_k\bi{\sf v}_k, \bi{\sf H}_k^2\bi{\sf v}_k, \dots, 
\bi{\sf H}_k^{n_a}\bi{\sf v}_k\right\}. 
\end{equation}
The orthonormal set is formed using Gram-Schmidt orthogonalization.
Letting $\bi{\sf h}_k$ denote the $(n_a+1)\times (n_a+1)$ upper-Hessenberg matrix 
formed by the coefficients $(\bi{\sf h}_k)_{ij}$ we obtain the matrix equation
\begin{equation}
\bi{\sf h}_k = \bi{\sf Q}_k^\dagger \bi{\sf H}_k\bi{\sf Q}_k,
\end{equation}
where $\bi{\sf Q}_k$ is a matrix formed from the $n_a$ column vectors
$\left[\bi{\sf q}_0, \bi{\sf q}_1, \bi{\sf q}_2, \dots, \bi{\sf q}_{n_a}\right]$
and so $\bi{\sf h}_k$ is the Krylov subspace Hamiltonian which is 
calculated simultaneously with $\bi{\sf Q}_k$. 
$\tilde{\bi{\sf H}_k} = \bi{\sf Q}_k^\dagger \bi{\sf h}_k\bi{\sf Q}_k$ can be used as a 
replacement to $\bi{\sf H}_k$ in a wide variety of applications. In particular the 
time-evolution operator can be written as 
\begin{equation}
\tilde{\bi{\sf U}}(t_k+\Delta t, t_k) = e^{-i\tilde{\bi{\sf H}}_k\Delta t} = 
\bi{\sf Q}_ke^{-i\bi{\sf h}_k\Delta t}\bi{\sf Q}_k^\dagger.
\label{eqn:evolution_operator}
\end{equation}
Now $\bi{\sf h}_k$ is typically a tridiagonal matrix and so 
$e^{-i\bi{\sf h}_k\Delta t}$ can
be computed inexpensively. Thus $\tilde{\bi{\sf U}}(t_k+\Delta t, t_k)$ is in 
effect  a unitary propagator
correct to order $(\Delta t)^{n_a}$. 
\end{subsection}
\end{section}
\begin{section}{Accuracy of the discrete solutions}

The results we obtain from this method depend on several parameters, 
namely $N_\rho$, $h_\rho$, $\Delta z$, $z_{\max}$, $n_a$, $\Delta t$, and 
the choice of gauge used to describe the laser-molecule interaction 
term. In this section we detail the choices for these parameters 
required to obtain accurate and fully converged solutions.
\begin{subsection}{Spatial grid parameters}
In an intense field problem, the dynamics of the system can be very sensitive to the 
initial state. Therefore, the accuracy of the ground state wavefunction and its 
corresponding energy play a crucial role. 
Several important grid parameters are open to choice, and to 
determine these we proceed as follows. We first choose the values of $\Delta z$ and 
$N_{\rho}$, then apply a variational method by adjusting  the value of 
$0 \leq h_{\rho} \leq 2$ until we get an accurate ground-state energy. Importantly, 
once this
procedure is completed for a specific internuclear distance $R$, the same scaling factor 
works well for all $R$. In this hybrid finite-difference/discrete-variable method we 
noted that there is a delicate relation among the values of $\Delta z$, $N_{\rho}$ and 
$h_{\rho}$. For a  fixed $N_{\rho}$ value, if we double the value $\Delta z$, we have to 
roughly double the value of $h_{\rho}$ in order to maintain accurate  energies.
Since the kinetic energy terms are homogeneous in these coordinates this maintains 
the balance between these terms. In the following calculations, we take the range of the 
$z$ coordinate as $[-z_{\rm max},z_{\rm max}]$, where $z_{\rm max} = 300.9$ a.u. and
$\Delta z=0.1$ a.u.  ($6019$ points in all). The  $z$-subspace is shared across the 
processor array, in our case we used $13$ processors for this task. The $\rho$-subspace 
is spanned fully on each processor and we take  $N_{\rho}=30$ with  $h_{\rho}= 0.5185$ 
corresponding to the limit $\rho_{\rm max} \sim 55$ a.u. 
\begin{table}[t]
\caption{\label{groundstate} Calculated energies for the lowest states of gerade 
symmetry with $\Lambda=0$  and $\Lambda=1$ 
 for different $R$, compared with the exact values from Sharp's tabulation~\cite{Sharp}. 
 Values of $R$ and energy  are in atomic units. }
\begin{tabular}{@{}ccccc}
\hline
&\centre{2}{$\Lambda=0$ state }&\centre{2}{$\Lambda=1$ state }\\
 $R $   & Exact & Present    & Exact  &  Present \\   
\hline      
  1.0 & -0.451785 & -0.451783   &  0.525893 &   0.525872    \\
 2.0  & -0.602635& -0.602636   &  0.071229 &   0.071216 \\
 4.0  & -0.546085 & -0.546088  & -0.100825 &   -0.100830  \\    
 6.0  & -0.511968 & -0.511972  & -0.130325 &   -0.130327  \\
 8.0  & -0.502570& -0.502574   & -0.134511 &   -0.134512  \\
 10.0 & -0.500580 & -0.500582  & -0.132716 &   -0.132716  \\
 12.0 & -0.500167& -0.500172   & -0.129950 &   -0.129948  \\
 16.0 & -0.500035& -0.500040   & -0.126253 &   -0.126243   \\
 20.0 & -0.500015& -0.500018   & -0.125084 &   -0.125072   \\
\hline
\end{tabular}
\end{table}

To obtain the ground state eigenvectors it is convenient and efficient to use an 
iterative Lanczos method~\cite{Dundas2}. In Table~\ref{groundstate}, we list 
our ground state energies for $\Lambda=0$  and $\Lambda=1$ at different internuclear  
distances  and compare them with the exact values from Sharp's tabulation. Here, we 
have taken $\Delta z=0.1$ a.u. and $N_{\rho}=30$ in both calculation. However, the scaling 
parameter $h_{\rho}$ has been adjusted to $0.5185$ for $\Lambda=0$ and $0.146$ for $\Lambda=1$, 
these values are used for all $R$ in the table. The results are in excellent agreement
 compared with the exact results.

The excited 
state spectrum supported by the grid can be found using the spectrum of the 
autocorrelation function \cite{Dundas2} 
\begin{equation}
C(t) = \int d^3{\bi{r}} \phi^*({\bi{r}},t)\phi({\bi{r}},0), 
\end{equation}
where $\phi({\bi{r}},0)$ is an  arbitrary  trial function, and $\phi({\bi{r}},t)$ is 
the function evolved in the field-free Hamiltonian 
$\phi({\bi{r}},t) = \exp(-iH_0 t/\hbar)\phi({\bi{r}},0) $. The natural frequencies 
(eigenenergies) appear as peaks in the spectral density
\begin{equation}
    P(\omega,T) = \left\vert \int_0^{T} C(t)e^{i\omega t }dt
   \right\vert^2.
\end{equation}
The resolution improves with longer times, $T$.  
The trial function can be  chosen to find the states of given symmetry. 
We have compared the energies of the electronic states at an equilibrium 
separation  $R=2.0$ a.u. with the exact results given by
Sharp~\cite{Sharp}; this comparison was previously performed by 
Dundas~\cite{Dundas2}. Our current results are more accurate 
and better resolved than those given in \cite{Dundas2}. This has been achieved  
using a finer axial grid $\Delta z=0.1$ a.u. as compared with 
$\Delta z=0.2$ a.u.~\cite{Dundas2}, but also by adjustment of  
the scaling factor $h_{\rho}$. By trial and error, we determined that 
$h_{\rho} \approx 0.5185$ gave the best estimates over the full spectrum. The 
scaling factor can be considered a variational parameter~\cite{Baye1}. This is a 
useful and reliable method to determine the optimal grid parameters for subsequent 
dynamic calculations. 

At the very least, the dimensions of the cylindrical box, height $2z_{\rm max}$ radius
$\rho_{\rm max}$, must be chosen to encompass the tightly-bound states of the system. 
However, the evolution through  highly-excited diffuse states and low-energy continuum 
states is crucial to the ionization mechanism at low frequencies. The rescattering 
mechanism, by which  slow photoelectrons are driven back to the core region by the 
laser field, means that  the box size should be large enough to allow the continuum 
states to evolve unfettered. For example, the free-electron classical amplitude of 
displacement and momentum are proportional to $E_0/\omega_L^2$ and $E_0/\omega_L$, 
respectively and so box dimensions of several hundred atomic units may be required.

The boundary between bound and 
free electrons was established in two ways. We have defined an inner region which 
satisfies either
\begin{equation}
\sqrt{\rho^2 +(z-R/2)^2}\leq r_{\rm inner},
\end{equation}
or 
\begin{equation}
\sqrt{\rho^2 +(z+R/2)^2}\leq r_{\rm inner},
\end{equation}
where $R$ is the internuclear  separation distance and $r_{\rm inner}$ is taken 
to be $20$ a.u. We also define an outer box population to be the total 
population within the entire grid. 
   
In order to prevent reflection of ionizing wavefunction from the edges of the grid 
wavefunction splitting which acts in the same manner as an absorbing potential is applied 
in both the $z$ and $\rho$ directions~\cite{Dundas2}.
\end{subsection}
\begin{subsection}{Time propagation parameters}
\begin{figure}[t]
\centerline{\epsfclipon\epsfxsize=8.0cm\epsffile{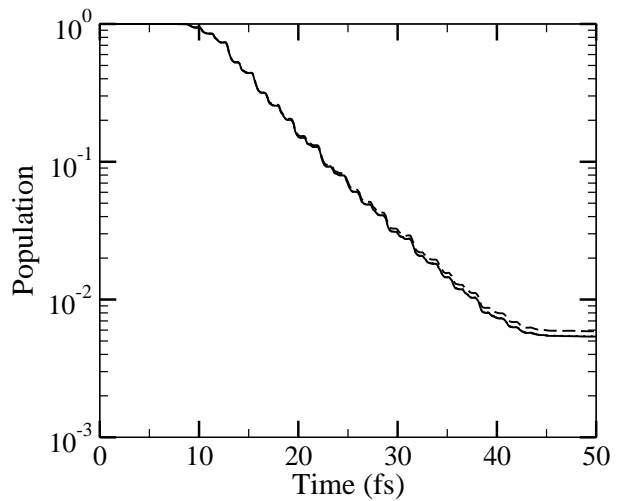}}
\caption{Population loss for different 
time steps $\Delta t$. The characteristics of the laser pulse are, $\lambda=800$ nm, 
$I =3.2 \times10^{14}$ W cm$^{-2}$, with  $\tau_1=5$ cycles (13.4fs)  and $\tau_2=10$ cycles 
(26.7 fs) and the bond length is $R=5$ a.u. The curves correspond to: $\full$ $\Delta t=0.02$ 
; $\dotted$$\Delta t=0.04$; $\dashed$$\Delta t=0.06$ in atomic units.} 
\label{timeconvg}
\end{figure}
A 12-th order Arnoldi propagator ($n_a$ = 12) was used in these
calculations. A test of the method for numerical stability with respect 
to the time step, $\Delta t$ using this choice of propagator order, $n_a$, in
conjunction with the spatial grid parameters outlined above is given in 
figure~\ref{timeconvg}. The logarithmic scale 
accentuates the differences in the residual population at the end of the 
pulse. In this case we note the very good convergence of the method.  
Results are presented for the following case: $\lambda=800$ nm and 
$I =3.2 \times10^{14}$ W cm$^{-2}$, and with  $\tau_1=5$ cycles (13.4 fs)  and 
$\tau_2=10$ cycles (26.7fs)  with  $R=5$ a.u. On this figure we note that only the 
largest time step $\Delta t = 0.06$ a.u. produces a small but discernible 
deviations in the results. We find that $\Delta t=0.05$ a.u. is sufficient for 
convergence in most cases, however the range $\Delta t =0.01-0.03$ a.u. 
provides more reliable and accurate results.
\end{subsection}
\begin{subsection}{Gauge invariance}
In Fig.~\ref{gauge800}, we show
the populations within the inner and outer boxes in both length gauge and velocity 
gauge. The bond length is $R=6$ a.u., and the laser pulse parameters are 
$\lambda=800 $ nm and $I =3.2\times10^{14}$ W cm$^{-2}$  with  $\tau_1=5$ cycles and 
$\tau_2=10$ cycles. Excellent agreement  between both gauges is observed. 
This agreement also extends to longer wavelengths and longer bond lengths.  
\end{subsection}
\end{section}
\begin{section}{Results}
\begin{subsection}{Static field ionization rates}
\begin{figure}[t]
\centerline{\epsfclipon\epsfxsize=8.0cm\epsffile{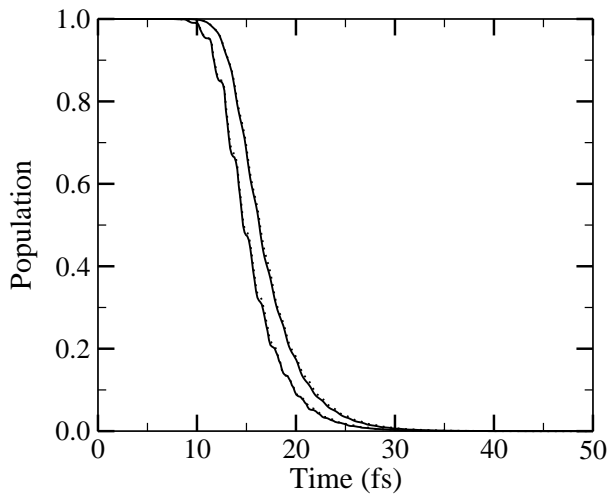}}
\caption{Electron population within the inner and outer boxes for length 
(full lines) and velocity (dotted lines) gauges.  
The two curves to the left correspond to the inner box, those to right, to 
the outer box. The bond length is $R=6$ a.u., and the laser pulse parameters are 
$\lambda=800 $ nm and $I =3.2\times10^{14}$ W cm$^{-2}$  with  $\tau_1=5$ cycles 
(13.4fs)   and $\tau_2=10$ cycles (26.7fs).} 
\label{gauge800}
\end{figure} 
The static field  ionization rates of the  hydrogen atom and  molecular ions 
are known to a high degree of accuracy using time-independent methods and 
 provide an important test of our method. In our time-dependent approach 
the field is switched on over a time $\tau_1=2$ fs, with the static field 
maintained at a constant value $F$ for $\tau_2=6$ fs. The rise in the field should be slow enough to 
ensure an adiabatic transition for the field-free ground state to the metastable 
state. In table~\ref{hatomdc}, we compare our results for the atom field ionization 
rates with the highly accurate time-independent results ~\cite{Plummer1, Plummer2}. 
Very good agreement is obtained in all cases.
\begin{table}[t]
\caption{\label{hatomdc} Static-field ionization rates $\Gamma$ for  the 
hydrogen atom. Comparison between time-independent (Floquet) calculations~\cite{Plummer2} 
and time-dependent methods (this work). The electric field strength $F$ is
given  in atomic units. The ionization rates are
quoted in the format $a(n)\equiv a\times10^n $ fs$^{-1}$.}
\begin{tabular}{@{}cccccc}
\hline
$F$~(a.u.)&0.1 &0.08 &0.06  &0.05338 &0.04  \\   
\hline
Floquet& 0.601(0) &0.188(0)  &0.213(-1)   &0.664(-2)   & 0.162(-3)  \\
Present& 0.600(0) &0.188(0) & 0.213(-1) & 0.664(-2) &0.163(-3)   \\
\hline
\end{tabular}
\end{table}
\begin{figure}[b]
\centerline{\epsfclipon\epsfxsize=8.0cm\epsffile{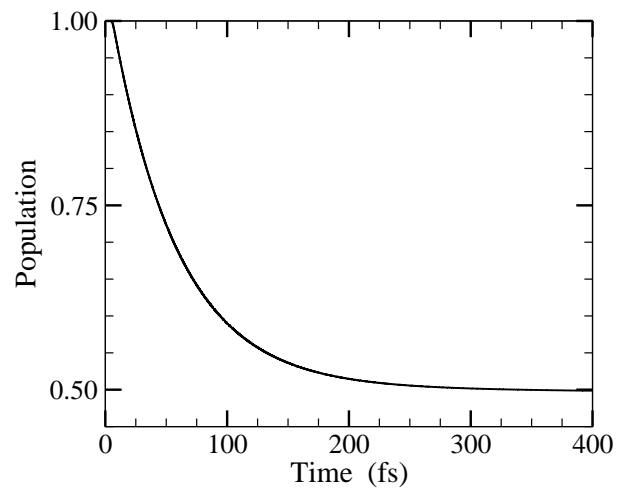}}
\caption{Electron bound-state population as a function of time for 
the H$_2^+$ ion. Static electric field strength $F=0.04$ a.u., internuclear distance 
$R=11$ a.u.. At longer times ($>200$fs) 
the decay is inhibited by a trapped state created during the 
rise of the electric field over a time $\tau_1=5$fs. }   
\label{poptrap}
\end{figure}

The same method can be applied to the molecular ion. However in this case the 
comparison is not so straightforward. Plummer and McCann \cite{Plummer1} noted that 
at large bond lengths the nearly degenerate pair of $\Sigma_{g,u}$  field-free states 
are strongly split by the external field. The correlated 
eigenstates are a pair of localised atomic resonances with 
large differences in their energies and their widths $\gamma_1,\gamma_2$. If we apply 
a static field over a time $\tau_1=5$ fs, relatively short in comparison to that 
characterising the gerade-ungerade splitting, that is the hopping time for the 
electron betweeen the centers, then the electron divides equally between 
the atom sites, creating an equal mixture of the resonance states rather than an 
adiabatic transfer to one or other state. Since $\gamma_1 \gg \gamma_2$ this would imply 
that half of the population is trapped. Consider 
the duration of the static field $T_s$ such that $\gamma_1 \ll T_s^{-1}$, then the 
population decay in the time-dependent model is given by 
$\Gamma = -dP/dt \approx {\textstyle \frac{1}{2}}   \gamma_1$, where $\gamma_1$ is the 
width of the short-lived state. If this were the case, then  at longer times $T_s \gg \gamma_1^{-1}$ but 
$T_s \ll \gamma_2^{-1}$ the population should reach a limit of 0.5. 
Figure~\ref{poptrap} shows  the population as a function of time with$\tau_1=5$ fs
and  $\tau_2=390$ fs for   $F=0.04$  and  $R=11$. The population decays gradually to 
a limiting value of 0.5 as predicted. However when the data is plotted on a logarithmic 
scale we notice that the decay process is not purely exponential. At the beginning, between $8$ fs 
and $13$ fs,  we estimate the ionization rate to be $8.75\times10^{-3}$ fs$^{-1}$ and 
near the end $3.91\times10^{-5}$ fs$^{-1}$ between $390$ fs and $395$ fs. The ionization 
rate calculated by the Floquet method is $1.76\times10^{-2}$ fs$^{-1}$ for the u-state 
and  $2.57\times10^{-5}$ fs$^{-1}$ for the g-state~\cite{Plummer1}. The factor of 2 
difference is consistent with the wavefunction splitting and population trapping. 
Further evidence is provided in figure~\ref{ionDCcomp}, with   ionization rates 
as a function of $R$, compared with those results obtained by Plummer and 
McCann~\cite{Plummer1} for the rate $\gamma_1$. At larger $R$ values the splitting is 
almost exactly one half in agreement with the adiabatic trapping model. However, at 
smaller values of $R$ we note that the degeneracy of the molecular states is removed 
and the ionization drops rapidly as the electron is able to move adiabatically into 
the trapped state. 
\begin{figure}[t]
\centerline{\epsfclipon\epsfxsize=8.0cm\epsffile{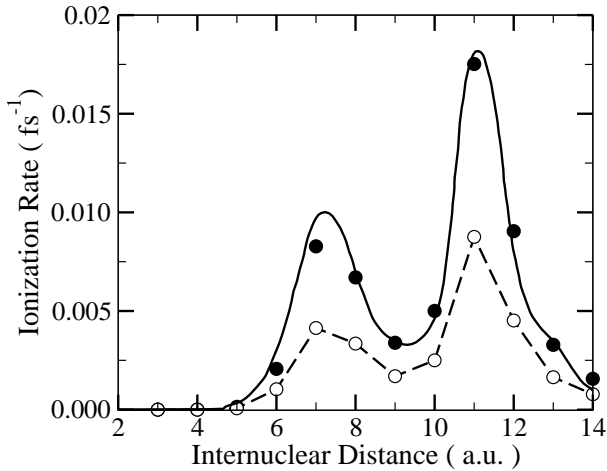}}
\caption{Ionization rate $\Gamma$  of H$_2^+$ as a function of internuclear 
distance $R$ in a static electric field  $F=0.04$ a.u. The results 
show that for the large bond lengths,  the electron splits 
evenly into a long-lived trapped state ($g$) and a 
short-lived resonance ($u$) : $\full$, Floquet calculation \cite{Plummer1} for the $u$ state; 
$\fullcircle$ present calculation $\times$ 2; $\opencircle$, present calculation.}   
\label{ionDCcomp}
\end{figure}
\end{subsection}
\begin{subsection}{Energy  shift of metastable states} 
The method can be applied to calculate the real part of the quasienergy, that is the 
Stark shift of the levels. In this case, we calculate the autocorrelation function 
$C(t)$ for a trial  function evolving with the external field on. We first calculate the shifts 
for the hydrogen atom and compare with other well-established results. In 
table~\ref{h_stark_shift}, results are compared with those obtained by the 
complex-coordinate approach~\cite{Maquet} for a static field DC Stark effect. Very good 
agreement is achieved. The resolution in the time-dependent method is limited by the 
bandwidth theorem: $\Delta \omega \approx 1/T_{p}$, where $T_{p}$ is the duration of 
the pulse.
\begin{table}[t]
\caption{\label{h_stark_shift} Stark shifts $\Delta$ for the hydrogen 
atom ground state. The electric field amplitude $F_0$  and laser angular 
frequencies are given in atomic units. $\omega_L=0$ corresponds to the 
static field. }
\begin{tabular}{@{}ccl}
\hline
 $ F_0$         &  0.10         &0.0354      \\   
$\omega_L$        & $0$          &0.375 \\ 
\hline      
Present results    &  -0.02746     & -0.0117  \\
Floquet method \cite{Maquet} & -0.02742        & -0.0119 \\
\hline
\end{tabular}
\label{tableiii}
\end{table}  
The calculation of the AC Stark shift is done in  the same way. We compare with the 
Floquet method applied to the hydrogen atom \cite{Maquet} for the angular frequency 
$\omega_L=0.375$. This is exactly resonant with the $2p_z$ state and thus, for moderate 
or low intensities, the AC Stark shift is to a good approximation half the Rabi 
frequency for the transition, that is
\begin{equation}
\Delta \approx - \frac{1}{2} F_0 \int d^3{\bi{r}} \phi_{1s}({\bi{r}})\ z \ \phi_{2p_z}({\bi{r}}) 
\end{equation} 
where $\phi_{1s}$ and $\phi_{2p_z}$ are the $1s$ and $2p_z$ wavefunctions respectively 
for the hydrogen atom, $F_0$ is the maximum of the electric 
field strength. For $F_0=0.0354$ a.u. then $\Delta \approx -0.0132$ a.u. A more precise 
estimate using  our code is $-0.0117$ a.u., which is in good agreement with the result of 
Maquet \emph{et al}~\cite{Maquet} $\Delta=-0.0119$ (see table \ref{tableiii}).
\begin{figure}[b]
\centerline{\epsfclipon\epsfxsize=8.0cm\epsffile{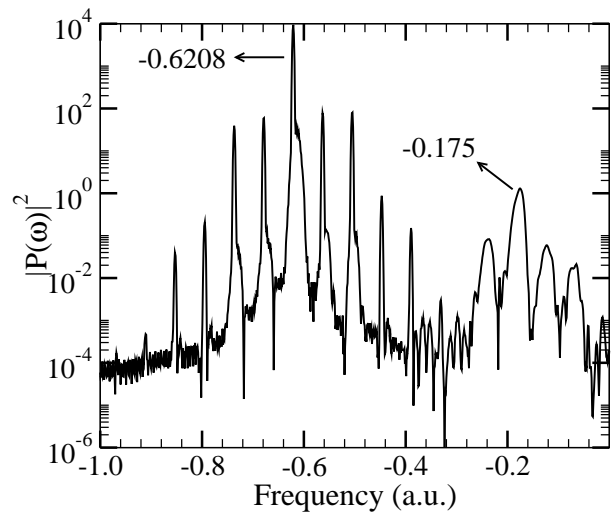}}
\caption{The quasienergies of H$_2^+$ in the presence of an intense laser field. The 
power-spectral density of the autocorrelation function is shown for a range of angular 
frequencies. In this case  
$\lambda = 800 $ nm, $I= 5\times10^{14}$ W cm$^{-2}$ and $R=2 $ a.u. .} 
\label{molestark2}
\end{figure}
Consider now the quasienergies of the molecular ions. In figure~\ref{molestark2} 
the spectral density for $R=2$ a.u. is given for the laser parameters 
$\lambda = 800$ nm, $I = 5\times10^{14}$ W cm$^{-2}$. In this case, the periodicity 
of the Floquet spectrum is clearly visible: $E_i+\Delta \pm n\omega_L$. The periodicity 
is extended and the peaks become sharper as the duration of the pulse increases. 
However, the resolution is limited by broadening due to the ionization process. The gap 
between any two neighbouring peaks is exactly one photon energy, i.e., $0.057$ a.u. 
for the present case. From the data we estimate the Stark shift for the ground state 
is $\Delta_g=-0.018$ a.u. and for the first excited state $\Delta_u=-0.0075$ a.u. 
The quasienergy spectrum results obtained using the  velocity-gauge are also shown in
figure~\ref{molestark1}. Both spectra have the same spectral line structure domain shown 
in the figure. The method is quite useful for calculation of isolated resonance shifts. 
However when overlapping resonances are present, for example at larger values of $R$ 
the resultant spectrum is very unclear and the method breaks down.

\begin{figure}[t]
\centerline{\epsfclipon\epsfxsize=8.0cm\epsffile{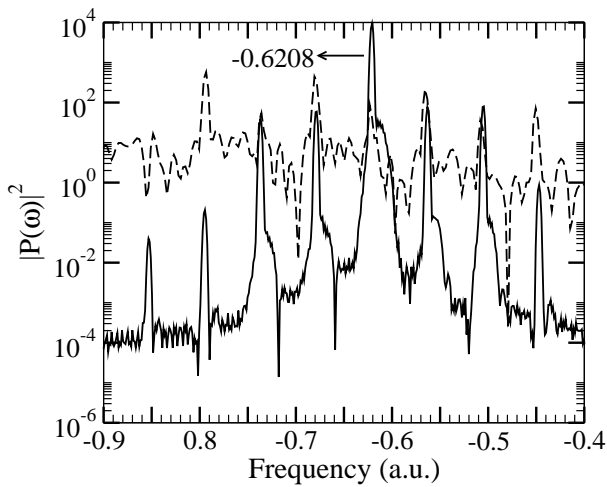}}
\caption{Spectral density of the correlation function showing the quasienergy spectrum. 
Comparison of the quasienergies of H$_2^+$ in the presence of intense laser field using 
different gauges. Here,$\lambda = 800 $ nm, $I= 5\times10^{14}$ W cm$^{-2}$ and 
$R=2 $ a.u..  $\full$, length gauge; $\dashed$, velocity gauge.} 
\label{molestark1}
\end{figure}
\end{subsection}
\begin{subsection}{Ionization of positronium}
The code can be easily adapted to the  ionization of positronium in an 
intense laser field. An investigation of this kind has been made by Madsen 
{\it{et al}}~\cite{Madsen} using a time-dependent basis-set expansion. The scaling factor 
$h_{\rho}$ is adjusted to be $0.52085$ for Ps with $\Delta z=0.1$ a.u. and 
$N_{\rho}=30$ to optimize the ground state energy to  $-0.250000001$ a.u.. 
 For the sake of comparison \cite{Madsen}, the velocity gauge is used and 
the pulse is taken as 
\begin{equation}
A(t) = A_0\sin^2\left(\frac{\pi t}{T_p}\right)\sin(\omega_L t),
\end{equation} 
where $A_0=F_0/\omega_L$ with $F_0$ the peak electric field strength. We take a pulse 
duration $T_p=50$ fs and a wavelength $\lambda= 780$ nm. The ionization probability 
versus laser intensity is shown in figure~\ref{positronium}. Our results are in very 
good agreement with those of Madsen et al~\cite{Madsen} over a wide range and  
especially for the intermediate intensities.
\begin{figure}[t]
\centerline{\epsfclipon\epsfxsize=8.0cm\epsffile{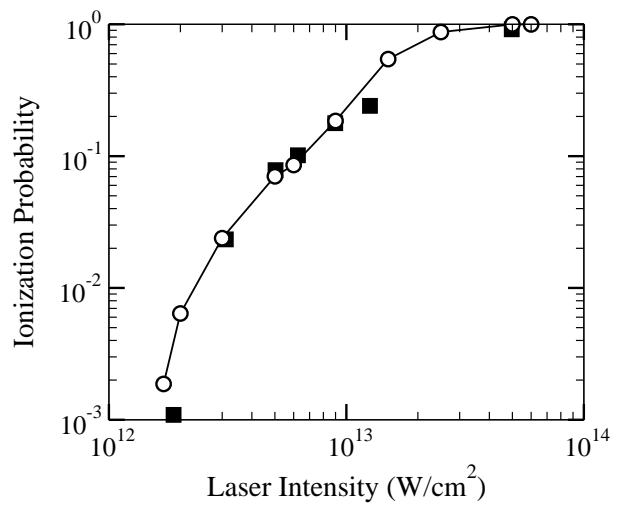}}
\caption{Ionization probability of Positronium at different laser intensities for 
wavelength $\lambda= 780$nm and pulse duration $T_p=50$ fs. $\opencircle$  Present 
Calculations; $\blacksquare$  Results of Madsen et al. ~\cite{Madsen}.} 
\label{positronium}
\end{figure}
\end{subsection}
\begin{subsection}{Ionization rates for H$_2^+$ by intense infrared light}
At very high intensities the bound states will ionize extremely quickly with a non-exponential decay. 
While the  ionization rate or width $\Gamma$ is mathematically well-defined,  one cannot 
calculate the rate so easily from observation of population decay.  In physical terms, in an experiment 
this corresponds to saturation, that is the molecules fully ionize before the pulse has finished. 
In this case it is difficult for experiments to analyse the response of the system. This requires 
extremely short pulses, with  associated rapidly-varying pulse envelope and broad bandwidth. 
Under such circumstance a time-dependent treatment is indispensable. 
The fragmentation of H$_2^+$  by an intense infrared laser ($\lambda = 790$ nm) 
is a problem of current  interest. For an intensity    
$I= 3\times$10$^{15}$ W cm$^{-2}$, and $R = 5$ a.u. the ionization rate is roughly  
$5.5$fs$^{-1}$. Thus the
molecule will be fully ionized within a fraction of an optical cycle.
Mathematically, the  ionization rates in this 
regime can be calculated  by  scaling as shown by Madsen et al. ~\cite{Madsen}. 
Using the arbitrary dimensionless parameters $\alpha$ and $\beta$ we can change the 
scale of length and time according to
\begin{equation}
\tilde{\bi{r}} = \alpha \beta \bi{r}, \ \ \ \ \ \ \ \ \tilde{t} = \alpha \beta^{2}  t,
\end{equation}
so that
\begin{equation}
\tilde{\omega} = \alpha^{-1} \beta^{-2} \omega_L.
\label{omegares}
\end{equation}

\begin{table}[t]
\caption{\label{scalingtest} Test of the scaling procedure. Ground-state energies 
and the corresponding ionization rates are calculated using scaled 
($\alpha=1.43$, $\beta=1.43$) and unscaled 
Hamiltonians. The laser parameters are 
$\lambda$=800 nm and I =3.2$\times$10$^{14}$ W cm$^{-2}$ and a range of 
bond lengths  $R$ are considered. Energy is in atomic units, ionization rate in  fs$^{-1}$ and 
 $R$   in   atomic units.}
\begin{tabular}{@{}ccccc}
\hline
 &\centre{2} {Scaled model} & \centre{2}{Unscaled model}  \\   
\hline
$R$ &Energy& Rate     & Energy &  Rate \\
\hline    
 4.0&-0.54600 & 0.128  &  -0.54608  &0.138 \\
 5.0&-0.52442  & 0.168  &  -0.52442  &0.165\\
 6.0&-0.51185 & 0.341  & -0.51199   &0.351 \\
 7.0&-0.50532 & 0.299  & -0.50559  &0.294 \\ 
 8.0&-0.50232 & 0.289  &-0.50257  &0.264 \\
 9.0&-0.50112 &  0.211      &  -0.50119  &0.196 \\
\hline 
\end{tabular}
\end{table}
The corresponding TDSE becomes 
\begin{eqnarray}
   i\frac{\partial }{\partial \tilde{t}} \psi(\tilde{\rho},\tilde{z},\tilde{t})
  = \Biggr[ & - &\frac{\alpha}{2\mu} 
    \left( \frac{\partial^2}
    {\partial \tilde{z}^2} +\frac{\partial^2} {\partial \tilde{\rho}^2} + \frac{1}{\tilde{\rho} }
   \frac{\partial}{\partial \tilde{\rho}}\right) \nonumber \\
   & +  & \tilde{V_e}(\tilde{R},\tilde{\rho},\tilde{z}  ) \nonumber \\
   &+&
   \tilde{V}_{l-m}(\tilde{z} ,\tilde{t}) \Biggr] \psi(\tilde{\rho},\tilde{z},\tilde{t}),  
 \label{shroedinger2}
\end{eqnarray}
with 
\begin{eqnarray}
 \tilde{V_e}(\tilde{R},\tilde{\rho},\tilde{z}  )  &=& 
 -\frac{Z_1}{\beta\sqrt{\tilde{\rho}^2 + (\tilde{z}-\tilde{R}/2)^2}}  \nonumber \\
 & - &\frac{Z_2}{\beta\sqrt{\tilde{\rho}^2 +(\tilde{z}+\tilde{R}/2)^2}} \nonumber \\
 & + &\!\!\frac{\alpha\Lambda^2}{2 \tilde{\rho}^2} +\frac{Z_1Z_2}{\beta \tilde{R}},
\end{eqnarray}
and  
\begin{equation}
 \tilde{V}_{l-m}(\tilde{z} ,\tilde{t})=  \alpha^{2}
 \beta^{3} \tilde{z} E(\tilde{\omega},\tilde{t}).
\end{equation} 
The laser intensity and the ionization rate scale as 
\begin{equation}
 \tilde{I} = \alpha^{-4} \beta^{-6} I,
\end{equation} 
and  
\begin{equation}
\tilde{\Gamma} = \alpha^{-1} \beta^{-2} \Gamma,  
\end{equation} 
accordingly.

\begin{figure}[t]
\centerline{\epsfclipon\epsfxsize=8.0cm\epsffile{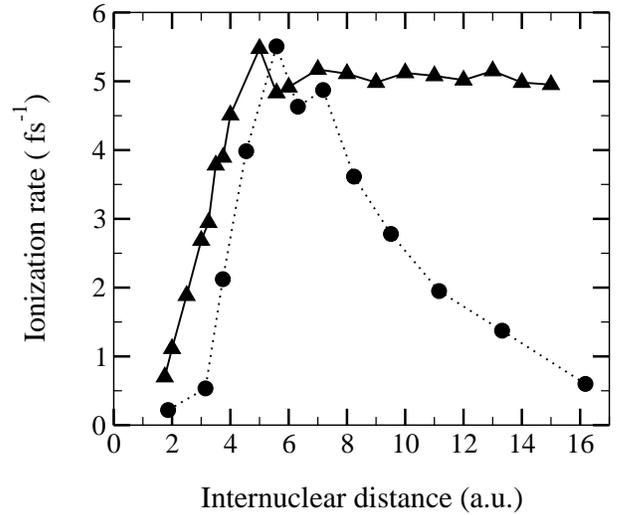}}
\caption{Ionization rates as a function of internuclear distance ($R$) with  
$\lambda= 790$ nm and $I = 3\times10^{15}$ W cm$^{-2}$. $\blacktriangle$,{\bf{ }} 
theoretical calculations using Hamiltonian rescaling procedure; $\fullcircle$, 
experimental measurements~\cite{Williams}. The experimental data
are normalized to the theoretical result at $R = 5$ a.u..} 
\label{highinten1}
\end{figure}
\begin{figure}[b]
\centerline{\epsfclipon\epsfxsize=8.0cm\epsffile{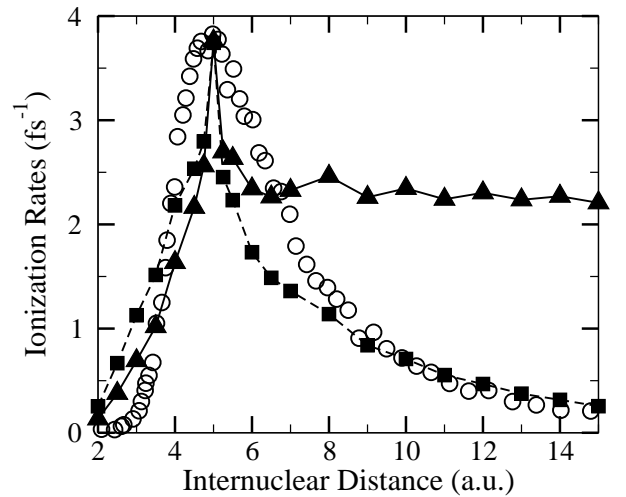}}
\caption{Ionization rates as a function of internuclear distance ($R$) with  
$\lambda= 800$ nm and $I = 1.4\times10^{15}$ W cm$^{-2}$.
 $\blacktriangle$, 
theoretical calculations using Hamiltonian rescaling procedure; $\fullsquare$, 
theoretical modelling with population depletion based on Coulomb explosion 
included, $dP/dR\approx -C (\Gamma/v) P$, where we take the constant $ C=0.2$;
$\opencircle$, experimental measurements~\cite{Gibson}. The experimental data 
are normalized to the theoretical result at $R = 5$ a.u..} 
\label{highinten2}
\end{figure}
The numerical stability scaling can be tested at intensities below saturation.
 In table~\ref{scalingtest},
we compare the scaled ground-state energies and ionization rates obtained with those 
from a direct calculation for $\lambda=800$ nm and $ I =3.2\times10^{14}$ W cm$^{-2}$ 
at various internuclear distances. The agreement is reasonable with relative errors in 
ionization rates below 9\%. The values of the scaling parameters in this case are 
$\alpha = 1.43 $ and $\beta = 1.43$. The interest in obtaining rates at high 
intensity follows experimental work  that estimated the bond length dependence of 
ionization rate from the ion energy spectrum~\cite{Gibson, Williams}. Our results 
are compared with experimental spectra in figures~\ref{highinten1} and~\ref{highinten2}  
taking $\alpha = 1.43$ and $\beta = 1.43$. The experimental measurements collect 
ions over a large part of the focal volume, and hence we adjust the normalization of the 
data to the theoretical results. It is surprising and remarkable to note that both spectra, 
experimental and theoretical, are dominated by a single large peak and that its location 
is reproduced accurately by the theory. For $R<7$ the shape of the ionization rate 
is quite well reproduced. This is surprising in view of the fact that the theoretical model is greatly simplified and 
does not include nuclear vibrations nor the averaging over molecular orientation and focal 
volume as would be required for a true comparison. One conclusion might be that the 
extremely good agreement indicates that the ionization rate is a strongly-peaked function 
of bond length, molecular orientation and laser intensity so that the averaging process 
does not broaden these features. For internuclear distance  greater than $7$ a.u., there 
is no dependence on bond length, indicating the loss of molecular effects. An essential 
assumption in the theoretical model is that the molecules are equally populated at all 
$R$ corresponding to the ionization process occurring as the molecule dissociates at 
steady speed. An explanation of the experimental shortfall in ions from large $R$ is that 
if the molecule ionizes fully at smaller $R$ it cannot survive to yield ions at large 
bond lengths and the ion yield rapidly drops. The other point in figure~\ref{highinten1} 
deserving note is the overall leftward displacement of our theoretical curve compared 
with the experimental one. There is the possibility that the experimental calibration 
of the laser intensity underestimates the actual intensity experienced by the molecules. 
A very rough simulation of the molecule depletion is shown in figure \ref{highinten2} is 
which the theoretical ion yield is exponentially attenuated. The curve shown is 
$P(R)= \Gamma(R) \exp( -C \int_{R_0}^R \Gamma(R)/v(R)\ dR )$ where $v(R)$ is the relative 
velocity of the ions taken to be the classical value for Coulomb repulsion
$m_p v^2/4 +1/R_0 =1/R$. The factor $C$ is an empirical constant taken to fit to the 
experimental curve in figure  \ref{highinten2}. We find $C\sim 0.2$ gives the 
best shape for the distribution. Taking $C=1$ leads to 
a severe loss of ions at large $R$, almost no ions survive beyond $R=7$ in this 
approximation. Of course one might expect that the low intensity focal averaging process 
might raise the yield of ions from large $R$ and give a more realistic picture of the 
process. This would require inclusion of the attenuation corrections discussed above. 
Nonetheless, it is fair to compare this theory and experiment for small $R$ values, and 
in this respect the agreement is remarkably good.
\end{subsection}
\end{section}
\vspace*{-0.5cm}
\begin{section}{Conclusions}
We have made a detailed investigation of a method which is designed to solve the 
reduced-dimensionality  time-dependent Schr\"{o}dinger equation for metastable 
systems in intense fields. We have checked the reliability 
of the present code by examining  the convergence and the gauge dependence. Applications  
to several problems have been carried out and yield good agreement  with other available 
theoretical results. However, by direct solution of the TDSE, our method can be applied 
to both short and long pulses and to a large variety of wavelengths. The provision 
of parallel computer architecture offers the opportunity to study such systems 
from first principles and in full dimensionality.
\end{section}
\begin{section}*{Acknowledgements}
LYP acknowledges the award of a PhD research studentship from
the International Research Centre for Experimental Physics (IRCEP) at Queen's
University Belfast. DD acknowledges the award of an EPSRC Postdoctoral
Fellowship in Theoretical Physics. This work has also been supported 
by a grant of computer resources at the Computer Services for Academic 
Research, University of Manchester and at HPCx, Daresbury Laboratory provided by 
EPSRC to the UK Multiphoton, Electron Collisions and BEC HPC Consortium.

\end{section}


\end{document}